\documentclass[useAMS,usenatbib]{mn2e}

\newcommand{\vunit}{\mbox{\,km\,s$^{-1}$}}
\newcommand{\mic}{\mbox{$\,\mu$m}}
\newcommand{\pion}[2]{{#1}\,{\sc {#2}}}
\newcommand{\fion}[2]{[{#1}\,{\sc {#2}}]}

\title[Dust in nova DZ Cru]{The peculiar dust shell of Nova DZ Cru (2003)} 
\author[A. Evans et al.]
{A. Evans$^{1}${\thanks{Email: {\tt ae@astro.keele.ac.uk}}},
R. D. Gehrz$^2$,
C. E. Woodward$^2$,
L. A. Helton$^2$,
M. T. Rushton$^3$, \newauthor
M. F. Bode$^4$,
J. Krautter$^5$,
J. Lyke$^6$,
D. K. Lynch$^7$,
J.-U. Ness$^8$,
S. Starrfield$^{9}$, \newauthor
J. W. Truran$^{10}$,
R. M. Wagner$^{11}$
\\
$^1$Astrophysics Group, Keele University, Keele, Staffordshire, ST5 5BG, UK\\
$^2$Department of Astronomy, School of Physics \& Astronomy, 116 Church Street
     SE, University of Minnesota, Minneapolis, MN 55455, USA\\
$^3$Centre for Astrophysics, University of Central Lancashire,
     Preston, PR1 2HE, UK \\
$^4$Astrophysics Research Institute, Liverpool John Moores University,
     Twelve Quays House, Birkenhead CH41 1LD, UK\\
$^5$Landessternwarte, Zentrum f\"ur Astronomie der Universit\"at Heidelberg,
Koenigstuhl, D-69117 Heidelberg, Germany \\  
$^6$W. M. Keck Observatory, 65-1120 Mamalahoa Highway, Kamuela, HI 96743, USA \\
$^7$The Aerospace Corporation, M2/266, P.O. Box 92957, Los Angeles, CA 90009,
    USA \\ 
$^8$European Space Astronomy Centre, PO Box 78, 28691 Villanueva de la Ca\~nada,
Madrid, Spain \\ 
$^9$School of Earth and Space Exploration, Arizona State University, PO
Box~871404, Tempe, AZ 85287-1404, USA\\
$^{10}$Joint Institute for Nuclear Astrophysics, The University of Chicago,
   Chicago, IL 60637, USA \\
$^{11}$Large Binocular Telescope Observatory, 933 North Cherry Avenue, Tucson, AZ
    85721, USA \\}

\begin{document}

\date{Version of 25/05/10}

\pagerange{\pageref{firstpage}--\pageref{lastpage}} \pubyear{2009}

\maketitle

\label{firstpage}

\begin{abstract}

We present {\it Spitzer Space Telescope} observations of the ``peculiar
variable'' DZ~Cru, identified by \citeauthor[][]{rushton} (2008, MNRAS, 386,
289) as a classical nova. A dust shell, on which are superimposed a number of
features, is prominent in the 5--35\mic\ range some 4~years after eruption. We
suggest that the dust in DZ~Cru is primarily hydrogenated amorphous carbon
in which aliphatic bands currently predominate, and which may either become
predominantly aromatic as the dust is photo-processed by ultraviolet
radiation from the stellar remnant, or more likely completely destroyed.
\end{abstract}


\section{Introduction}

DZ~Cru (also known as Nova Cru 2003 and the ``peculiar variable'' DZ Cru) was
discovered in 2003 August \citep{tabur}. It was initially reported as a nova but
spectroscopic observations gave rise to doubts about its nova status, and
comparisons were drawn with the unusual variable V838~Mon \citep{dellavalle}.
\cite{rushton} presented  1--2.5\mic\ infrared (IR) spectra obtained
$\sim1-1.5$~years after discovery. They found \pion{H}{i}, \pion{O}{i} and
\fion{N}{i} emission lines superimposed on a hot dust continuum and concluded
that DZ~Cru was indeed a classical nova 
at distance $\sim9$~kpc (although this is subject to considerable uncertainty).

Here we present observations of DZ~Cru with the {\it Spitzer} Space Telescope
\citep{spitzer,gehrz-spitzer}, that reveal a hot dust shell with unusual
emission features. 

\section{Observations}

DZ~Cru was observed with the InfrarRed Spectrograph \citep[IRS;][]{houck} on {\it
Spitzer} in staring mode on three occasions, as detailed in Table~\ref{obs}.
Spectra were obtained with both low- and high-resolution modes, covering the
spectral range of 5--38\mic, and the blue peak-up array was used to centre the
object in the IRS slits. For the high-resolution modes we also obtained
observations of the background. The spectrum was extracted from the version 12.3
processed pipeline data product using {\sc spice} version 2.2 \citep{spice}.

We discuss primarily the low-resolution data in this paper, and use the
high-resolution data (which will be presented in detail elsewhere) to determine
expansion velocities; the gas-phase features will also be discussed elsewhere. The
spectra are shown in Fig.~\ref{dzcru_dust}.

\begin{table*}
 \centering
\begin{minipage}{140mm}
  \caption{{\it Spitzer} observational log. \label{obs}}
  \begin{tabular}{cccccc}
  \hline
UT at mid-observation  & Programme & AOR & Time from     &  Observing & Dust flux\\
(YYYY/MM/DD.DD) &        & Key  &  outburst (d)  &  time (s) & ($10^{-13}$~W~m$^{-2}$) \\ \hline
2007/09/05.86   & 30076   &  22269952  &  1477          & 2692 & 9.8\\
2008/02/25.46  & 40060   & 17734912 &  1650          &  698 & 7.9\\
2008/04/21.33   & 30076   & 17735168 &  1706          & 1783 & 7.7 \\ \hline
\end{tabular}
\end{minipage}
\end{table*}

\section{Discussion}

The spectral energy distributions (SED) are clearly dominated by dust, with
emission features due to \fion{Ne}{iii} (15.5\mic) and \fion{O}{iv} (25.9\mic),
and dust emission features, superimposed; \fion{Ne}{ii} at 12.8\mic\ is not
present, to a limit of $2.7\times10^{-18}$~W~m$^{-2}$. We discuss the deduced
expansion velocities, the dust emission, and the dust features in turn.

\subsection{Expansion velocities}

The FWHM of the (gas-phase) emission features indicate a velocity of
$800-1000$\vunit, assuming Doppler broadening of the lines. The {\it Spitzer}
derived values are broadly  in agreement with the
velocities reported by \cite{rushton} on the basis of line widths in the
1--2.5\mic\ range. The presence of \fion{Ne}{iii}~15.5\mic\ and
\fion{O}{iv}~25.9\mic\ emission is common in mature novae \citep{lynch,schwarz};
this, together with the inferred expansion velocities, appear to reinforce the
conclusion of \cite{rushton} that DZ~Cru is a classical nova.

If we assume that the \pion{O}{i} and \fion{O}{iv} lines arise in the same
region of the ejecta (and given the time interval between the ground-based and
{\it Spitzer} data, this assumption may be not be valid), we can assess whether
the ejecta speed has changed over the $\sim1000$~day period covered by our data.
The velocities (deconvolved for instrumental resolution) derived from the
\pion{O}{i}~1.317\mic\ line \citep{rushton} and the \fion{O}{iv}~25.9\mic\
line in our high-resolution {\it Spitzer} data over the $\sim 4.5$~yr period of
our data satisfy
\begin{equation} V_{\rm exp} = 1480 \, (\pm300) - 0.37\, (\pm0.23)  \,\,
t\rm{(days)}  \:\: ,\label{Vexp}
\end{equation}
which indicates no significant variation with time. Thus we adopt the assumption
that the velocity is constant, and we take the mean value $\overline{V}_{\rm
exp} = 1030\pm110$\vunit\ for the period covered by our observations.

\begin{figure*}
\setlength{\unitlength}{1cm}
\begin{center}
\leavevmode
\begin{picture}(5.0,8.5)
\put(0.0,4.0){\includegraphics{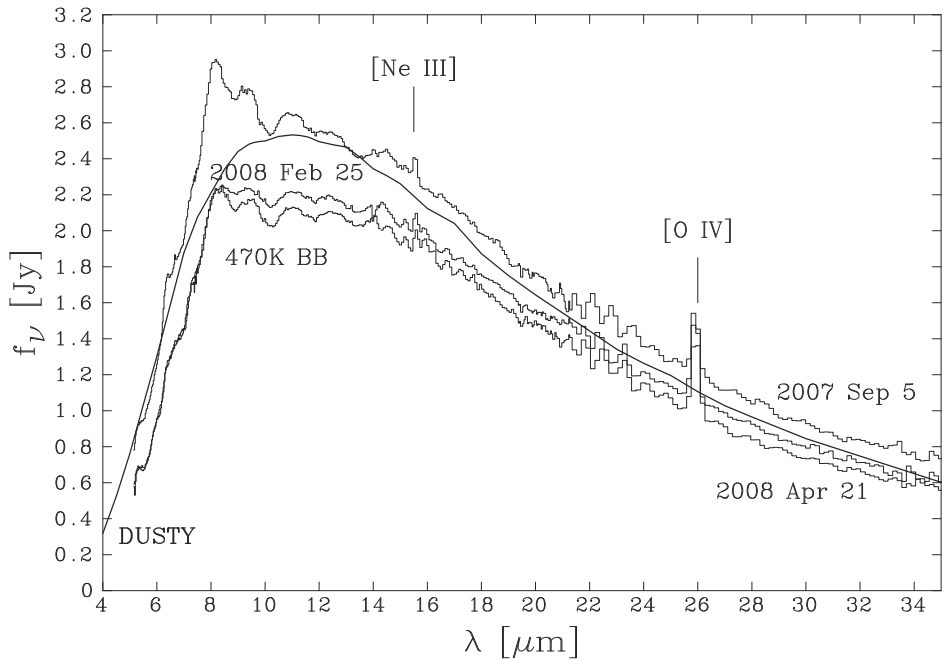}}
\end{picture}
\caption[]{The dust shell of DZ~Cru on the dates shown. The curve labelled
``{\sc dusty}'' (red) has been normalized to the 2007 September spectrum at 10.1\mic.
Gas phase emission features are identified. The ``fringes'' in the 20--24\mic\
region are instrumental artefacts.} 
\label{dzcru_dust} 
\end{center}
\end{figure*}

\subsection{Dust temperature and flux evolution}
\label{s_temp}

We have limited information about the state of the stellar remnant and of
the ejecta at the time of our {\it Spitzer} observations. However it is evident
that the SED of the dust is not well represented by a black-body, and we fit the
2007 September data using {\sc dusty} \citep{dusty}. We suppose that the stellar
remnant is a black-body with temperature $10^5$~K, the emitting dust is amorphous
carbon with optical constants from \cite{hanner}, the dust shell consists of
0.2\mic\ grains, has an $r^{-2}$ density distribution and an optical depth at
0.55\mic\ of 0.1 \cite[see][for the rationale for these values, to which the
output is not highly sensitive]{evans-iv}.

The temperature at the inner radius of the dust shell of the {\sc dusty} model
that best fits (by eye) is 450~K; the fit, normalized to the dust emission for
2007 September (day~1477) at 10.1\mic, is included in Fig.~\ref{dzcru_dust}.
Assuming an absolute bolometric magnitude of $-8.3$ for DZ~Cru \citep{rushton},
the output from {\sc dusty} indicates that the inner radius is at distance
$\sim7.5\times10^{13}$~m from the stellar remnant. This implies that the
dust-bearing material was carried out at velocity $\sim580$\vunit, significantly
slower than the oxygen-bearing material (Equation~(\ref{Vexp})), but comparable
with velocities reported ($\sim500$\vunit) by \cite{dellavalle} about 2~days
after discovery.

The flux from the dust shell, $f$, integrated over the wavelength range
5--35\mic, is given in Table~\ref{obs}, although as we have no data outside this
range the values given are lower limits; nonetheless the flux clearly declines
over the time of our {\it Spitzer} observations (see also
Fig.~\ref{dzcru_dust}). For a free-flowing expansion, constant $L_*$ and
conservation of grains, we expect $|\delta{f}/f|\sim2\,\delta{t}/t$. Within the
uncertainties, the decline in the dust flux (Table~\ref{obs}) is consistent with
this expectation and, therefore, with the assumption of constant expansion
velocity. We also note that \cite{rushton} estimated dust temperature
$690\pm40$~K ($620\pm50$~K) in 2005 February (June); the decline in temperature
from 2005 to the {\it Spitzer} observations in 2007-8 are also consistent,
within the uncertainties (and further noting that the dust temperatures in
\citeauthor{rushton} are not black-body), with a free-flowing expansion. 

It is not straightforward to compare the dust flux with the outburst flux as
both luminosity at maximum and distance are poorly known. Using the values for
absolute magnitude at maximum ($-8.3$) and distance (9~kpc) from \cite{rushton} we
estimate that the un-extinguished nova flux at outburst was
$\sim6.5\times10^{-11}$~W~m$^{-2}$, some two orders of magnitude higher than our 
estimated dust fluxes in Table~\ref{obs}.

The length of time for which strong emission by the dust shell has persisted is
unusual for a classical nova: the dust is still  emitting strongly after
$\sim4.5$~years. By contrast, in the case of the archetypical dusty nova
V705~Cas, the dust shell -- prominent in the first two years after eruption
\citep{mason,evans-ii,evans-iv} -- was not detected with the Infrared Space
Observatory ISO some 950, 1265 and 1455~days ($\sim4$~years) after outburst, to
$3\sigma$ limits of 0.8~Jy in the wavelength range 5--12\mic\ \citep{salama}.

One possible explanation might be that the material ejected in the 2003 eruption
has decelerated, ``snow-ploughing'' into circumstellar material that pre-dated the
eruption. Such a scenario might be plausible if there were evidence for a 
decrease in ejecta velocity but as already noted (see Equation~(\ref{Vexp})),
there is no such evidence. In addition, interaction between fast-moving ejecta and
dense circumstellar material would lead to shocked emission in the radio and X-ray
bands, as in the case of GK~Per \citep{gkper} and coronal emission in the IR.
There is no evidence for either effect. Furthermore there is no evidence for any
emission from the position of the progenitor in the 2MASS, MSX and IRAS surveys.

\subsection{Dust emission features}

\begin{table*}
 \centering
  \caption{Dust features in DZ~Cru in 2007 September; an ``S'' means that the
  feature is a   ``shoulder'' on a main peak, ``P''. Presence ($\sqrt{}$) or
  absence ($\times$) in the spectra of the novae V705~Cas and V2362~Cyg, and the post-AGB
  objects CRL2688 \citep{peeters} and HD100764 \citep{sloan}, are also indicated; a $-$
  indicates no information.     \label{features_id}} 
  \begin{tabular}{ccccccccc}
$\lambda$ (\mic) & FWHM (\mic) & S/P & V705~Cas &  V2362~Cyg & CRL\,2688 &
HD\,100764 & ID \\
                 &             &     &          & & & & \\ \hline

$6.46\pm0.01$    & $0.58\pm0.03$ & P &   -- & $\sqrt{}$ & $\sqrt{}$ ($\sim6.3$)& $\sqrt{}$ (6.33) & UIR\\
$7.24\pm0.01$    & $0.39\pm0.02$ & S &   -- & -- & $\times$ & $\times$ & --\\
$8.12\pm0.01$    & $0.97\pm0.02$ & S & $\sqrt{}$ & $\sqrt{}$ & $\sqrt{}$ (8.22) & $\sqrt{}$ (8.15) & UIR \\
$9.29\pm0.01$    & $0.94\pm0.02$ & P &  $\times$  & $\times$ & $\times$ & $\times$ & -- \\ 
$10.97\pm0.03$   & $0.93\pm0.06$ & P &  $\times$  & $\times$ & $\times$ & $\times$ & -- \\
$12.36\pm0.03$   & $0.88\pm0.07$ & P &   $\times$ & $\times$ & $\times$ & $\times$ & -- \\ \hline
\end{tabular}
\end{table*}

\begin{figure}
\setlength{\unitlength}{1cm}
\begin{center}
\leavevmode
\begin{picture}(5.0,6.)
\put(0.0,4.0){\includegraphics{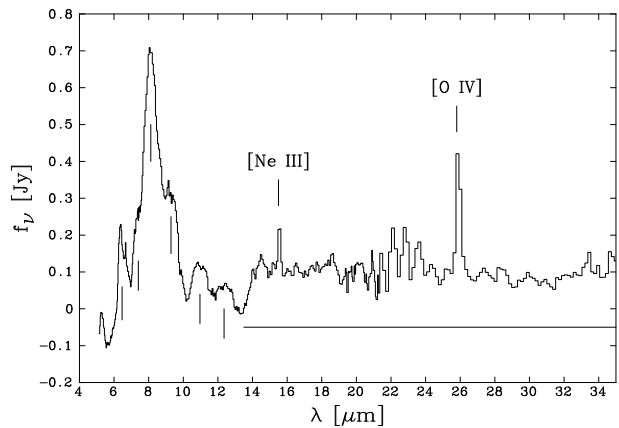}}
\end{picture}
\caption[]{Dust features, obtained by subtracting the {\sc dusty model} from the
2007 September spectrum. Central wavelength and FWHM of features identified by 
vertical tick-marks are given in Table~\ref{features_id}; the horizontal line
indicate the approximate extent of the ``plateau'' feature.}
\label{features} 
\end{center}
\end{figure}

There are superficial similarities between the dust shell of DZ~Cru and those of
V2361~Cyg (Helton et al., 2010, in preparation) and V2362~Cyg \citep[][Helton
et al., 2010, in preparation]{lynch}.  The latter displayed features at 6.37,
8.05, 11.32, and a broad ``plateau'' centred at $\sim18$\mic; \citeauthor{lynch}
were unable to securely identify any of the dust features in V2362~Cyg.

We use the {\sc dusty} fit (see Section~\ref{s_temp} and Fig.~\ref{dzcru_dust}) as a
baseline to extract the dust features that are clearly present in DZ~Cru. The result
is shown in Fig.~\ref{features}, in which the features are indicated by vertical
ticks. The wavelengths and FWHM of the features are given in
Table~\ref{features_id}, together with an indication of their presence (possibly at
slightly different wavelengths) or absence in V705~Cas \citep{evans-iv} and
V2362~Cyg \citep{lynch}, and in evolved stars. In addition there may be a
``plateau'' of emission that extends from $\sim14$\mic\ to $\sim34$\mic\ (cf.
Fig.~\ref{features}), although the shape of this depends of course on the shape of
the fitted {\sc dusty} spectrum longward of 14\mic. As is evident from
Fig.~\ref{dzcru_dust}, these features are present on all three dates of
observation, but we note that the 8.2\mic\ feature in DZ~Cru appeared to weaken in
the 2008 February spectrum, only to recover somewhat by 2008 April. The features
listed in Table~\ref{features_id} are significantly broader than the \fion{O}{iv}
and \fion{Ne}{iii} lines ($\mbox{FWHM}\sim0.5-1$\mic, compared with
$\sim0.15$\mic\ for the ionic lines), so it is unlikely that the features in
Table~\ref{features_id} are ionic lines, or even blends.

We note that neither the 9.7\mic\ nor the 18\mic\ silicate features are present
in the dust emission; furthermore, there is no evidence for the more structured
9.7\mic\ feature sometimes seen in cometary silicates \citep{wooden}. We are
therefore confident in ruling out silicates as a major dust component in DZ~Cru.

Also, it seems unlikely that any of the features in Table~\ref{features_id} can
be associated with silicon carbide, which displays a number of features in the
9--12.5\mic\ range, depending on the nature of the SiC \citep{speck}. No set of
features for the materials studied by \citeauthor{speck} match those listed in
Table~\ref{features_id}.

\section{UIR features}

Despite the absence of the expected 11.25\mic\ feature (but see below), some of
the features in DZ~Cru may plausibly be identified with ``Unidentified
infrared'' \citep[UIR; see e.g.][]{tielens} features, which are seen in
classical novae with optically thick dust shells
\citep{mason,evans-iv,ER,gehrz}. DZ~Cru is the fourth dusty nova in
which the ``8.2-\mic'' feature has been prominent \citep[][ Helton et al., 2010,
in preparation]{mason,evans-iv,lynch}. In the case of V705~Cas,
\cite{evans-ii,evans-iv} identified the prominent feature at 8.2\mic\ (cf.
Fig.~\ref{features}) as the UIR feature normally seen at 7.7\mic\ (see also
\citealt{geballe,mason}).

\cite{geballe} classified the UIR features according to the strength of features in the 3\mic\
and 6--13\mic\ regions; novae seemed to occupy a unique category, with a
3.4\mic\ feature that is strong relative to that at 3.28\mic.
Further classification of the UIR features has been carried out by
\cite{peeters} and from the point of view of this paper, the classification of
interest is their Class~{C}. Objects in this class show no 7.7\mic\ feature but
instead a feature at 8.2\mic; moreover they also display a weak 6.4\mic\
feature, and an extremely weak 11.25\mic\ feature. These characteristics are very
reminiscent of those seen in DZ~Cru (cf. Table~\ref{features}), and it is
tempting to suggest that the UIR features in DZ~Cru place it in
\citeauthor{peeters}'s UIR Class~C.

The two objects in \citeauthor{peeters}'s survey in their Class~C are
IRAS\,13416--6243 and CRL\,2688 (the ``Egg'' Nebula), both of which are post-AGB
stars, the latter undergoing rapid evolution from the AGB to the planetary
nebula phase \citep{cox}. An object with a similar suite of UIR features is
HD\,100764, a first-ascent red giant with a dusty disc \citep{sloan}. The
central star of CRL\,2688 has spectral classification F5Iae (effective
temperature $T_{\rm eff} \simeq6370$~K), while that of HD\,100764 is an early
carbon star with $T_{\rm eff}\simeq4850$~K; the $T_{\rm eff}$ of
IRAS\,13416--6243 is 5440~K \citep[][]{peeters,sloan}. Other objects displaying
similar UIR features have effective temperatures of this order \citep{sloan}.

Investigation of UIR features in evolved and Herbig Ae/Be stars
\citep{peeters,sloan,keller} strongly indicate that the wavelength of the
``7.7'' feature is highly environment-dependent, with a clear dependence of
wavelength on the effective temperature of the exciting star
\citep{sloan,keller}. In stars having {\em lower} effective temperatures (e.g.,
as evidenced by direct spectroscopy, or by lack of circumstellar ionization) the
``7.7'' feature has a systematically {\em longer} wavelength; for example the
``7.7'' feature only appears at 7.7\mic\ for $T_{\rm eff} \ga 10^4$~K.
\cite{sloan} and \cite{keller} attribute this to photo-processing of the UIR
carrier by the stellar radiation field, and the consequent shift from aliphatic
to aromatic bonds in the UIR carrier, which in novae is most likely a form of
hydrogenated amorphous carbon (HAC; we note that, while \cite{peeters} attribute
the carrier to PAH molecules, this is not the case in classical novae
\citep{ER1}).

If DZ~Cru is a classical nova, its effective temperature
$\sim4.5$~years post-eruption would be well in excess of the values for AGB
stars \citep[cf. the estimated $T_{\rm eff}\sim10^5$~K for V705~Cas
in][]{evans-iv}. There is ample evidence (e.g. from high excitation
emission lines) for a strong ultraviolet radiation field in all four novae
displaying the ``8.2'' feature. DZ~Cru, and the other novae in which the ``8.2''
feature have been present, seem to defy the trend followed by other UIR-bearing
objects.  Furthermore, the UIR feature that normally appears at 11.25\mic\
appeared at 11.4\mic\ in V705~Cas \citep{evans-iv}, which is also apparently
contrary to the trend of increasing wavelength with decreasing $T_{\rm eff}$ for
the ``11.25'' feature \citep{keller}.

The fact that the UIR features in DZ~Cru and
other novae are more akin to those in objects with a weak radiation field and
late spectral type suggest that the distinctiveness of UIR features in novae is
unconnected with the radiation environment. A likely explanation is that
the UIR carrier in DZ~Cru has only recently been exposed to the hard radiation
field of the stellar remnant, possibly because it has been protected within
dense clumps. Such clumps are required in nova winds to enable the chemistry that
leads to nucleation \citep{ER}, so that the (hydrocarbon) dust they contain
remains unaffected by the hard radiation field; there is certainly evidence for
continued clumpiness in the (optical) remnants of the dusty novae FH~Ser and
RR~Pic decades after eruption \citep{GOB1,GOB2}. The UIR carrier in DZ~Cru may
therefore have only recently been exposed to the ultraviolet radiation field of the
nova, thus limiting the photo-processing it has experienced.

\section{Concluding remarks}

We have described the dust shell of the classical nova DZ~Cru, which shows
several broad emission features. A number of these features remain unidentified.
However the ``8.2'' feature in DZ~Cru and other novae may be associated with a
UIR carrier in which aliphatic bonds predominate, likely as a consequence of the 
continued presence of dense clumps in the ejecta which shield the carrier
from the hard radiation field of the nova. We anticipate that, as the ejecta (and
clumps) disperse, a shift from aliphatic to aromatic may occur,
accompanied by the shift of the UIR features to the blue; however it is
more likely that the carrier will not survive exposure to the radiation. There
are also implications for features in the 3\mic-band, and it is unfortunate that
spectra in this crucial region do not appear to have been obtained in the case
of DZ~Cru.

Continued IR spectroscopy of DZ~Cru in the 3\mic\ and 7--13\mic\ regions would
be valuable to observe the effects of dispersal on the UIR (and other)
features. 

\section*{ACKNOWLEDGMENTS}

We thank the referee, Greg Sloan, for his thorough reading of an earlier
version of this paper, and for his constructive comments.

This work was supported in part by NASA/JPL {\it Spitzer} grants 1289430 and
1314757 to the University of Minnesota. SS acknowledges partial support from
NASA and NSF grants to ASU.

This research has made use of the SIMBAD database, operated at CDS, Strasbourg,
France.


\bsp

\label{lastpage}


\begin{thebibliography}{99}

	 
\bibitem[\protect\citeauthoryear{Cox et al.}{1996}]{cox} Cox, P., et al., 1996, A\&A,
         315, L265

\bibitem[\protect\citeauthoryear{Della Valle et al.}{2003}]{dellavalle} Della
         Valle, M., Hutsmekers, D., Savianne, I., Wenderoth, E., 2003, IAUC~8185

\bibitem[\protect\citeauthoryear{Evans et al.}{1997}]{evans-ii} Evans, A., Geballe,
        T. R., Rawlings, J. M. C., Eyres, S. P. S., Davies, J. K., 1998, MNRAS,
	292, 192

\bibitem[\protect\citeauthoryear{Evans \& Rawlings}{1994}]{ER1} Evans, A.,
         Rawlings, J. M. C., 1994, MNRAS, 269, 427

\bibitem[\protect\citeauthoryear{Evans \& Rawlings}{2008}]{ER} Evans, A., Rawlings,
         J. M. C., 2008, in {\it Classical Novae}, p.~308, second edition, eds M. F. Bode, A.
	 Evans, Cambridge University Press

\bibitem[\protect\citeauthoryear{Evans et al.}{2005}]{evans-iv} Evans, A., Tyne, V. H.,
        Smith, O., Geballe, T. R., Rawlings, J. M. C., Eyres, S. P. S., 2005, MNRAS,
	360, 1483

\bibitem[\protect\citeauthoryear{Geballe}{1997}]{geballe} Geballe, T. R., in
        {\it From Stardust to Planetisimals}, ASP Conference Series, Vol.~122,
	eds Y. J. Pendleton, A. G. G. M. Tielens, p.~119

\bibitem[\protect\citeauthoryear{Gehrz}{2008}]{gehrz} Gehrz, R.D., 2008, in {\it
         Classical Novae}, p.~167, second edition, eds M. F. Bode, A. 
	 Evans, Cambridge University Press


\bibitem[\protect\citeauthoryear{Gehrz et al.}{2007}]{gehrz-spitzer} Gehrz, R.
         D., et al., 2007, Rev. Sci. Inst., 78, 011302

\bibitem[\protect\citeauthoryear{Gill \& O'Brien}{1998}]{GOB1} Gill, C. D.,
         O'Brien, T. J., 1998, MNRAS, 300, 221

\bibitem[\protect\citeauthoryear{Gill \& O'Brien}{2000}]{GOB2} Gill, C. D.,
         O'Brien, T. J., 1998, MNRAS, 314, 175
	 

\bibitem[\protect\citeauthoryear{Hanner}{1988}]{hanner} Hanner, M., 1988, 
        NASA Conference Publications 3004, p.~22
	
\bibitem[\protect\citeauthoryear{Houck et al.}{2004}]{houck} Houck J. R., et al.,
        2004, ApJS, 154, 18 

\bibitem[\protect\citeauthoryear{Ivezi\'c \& Elitzur}{1997}]{dusty} Ivezi\'c,
        \v{Z}, Elitzur, M., 1997, MNRAS, 287, 799

\bibitem[\protect\citeauthoryear{Keller et al.}{2008}]{keller} Keller, L. D., et
         al., 2008, ApJ, 684, 411


\bibitem[\protect\citeauthoryear{Lynch et al.}{2008}]{lynch} Lynch, D. K., et
         al., 2008, AJ, 136, 1815

\bibitem[\protect\citeauthoryear{Mason et al.}{1998}]{mason} Mason, C. G., Gehrz, R.
         D., Woodward, C. E., Smilowitz, J. B., Hayward, T. L., Houck, J. R., 1998,
	 ApJ, 494, 783

\bibitem[\protect\citeauthoryear{Peeters et al.}{2002}]{peeters} Peeters, E.,
         Hony, S., Van Kerckhoven, C., Tielens, A. G. G. M., Allamandola, L. J.,
	 Hudgins, D. M., Bauschlicher, C. W., 2002, A\&A, 390, 1089

\bibitem[\protect\citeauthoryear{Rushton et al.}{2008}]{rushton} Rushton, M. T.,
         Evans, A., Eyres, S. P. S., van Loon, J. Th., Smalley, B., 2008, MNRAS,
	 386, 289

\bibitem[\protect\citeauthoryear{Salama et al.}{1999}]{salama}  Salama, A., Eyres, S.
         P. S., Evans, A., Geballe, T. R., Rawlings, J. M. C., 1999, MNRAS, 304, L20

\bibitem[\protect\citeauthoryear{Schwarz et al.}{2008}]{schwarz} Schwarz et al.
        2008, AJ, 134, 516 

\bibitem[\protect\citeauthoryear{Seaquist et al.}{1989}]{gkper} Seaquist, E. R.,
         Bode, M. F., Frail, D. A., Roberts, J. A., Evans, A., Albinson, J. S.,
	 1989, ApJ, 344, 805


\bibitem[\protect\citeauthoryear{Sloan et al.}{2007}]{sloan} Sloan, G. C., et al., 2007, ApJ,
         664, 1144

\bibitem[\protect\citeauthoryear{Speck, Thompson \& Hofmeister}{2005}]{speck}
         Speck, A. K., Thompson, G. D., Hofmeister, A. M., 2005, ApJ, 634, 426

\bibitem[\protect\citeauthoryear{Spice}{2009}]{spice} Spice User Guide, 2009,
         version GUI~2.2, {http://ssc.spitzer.caltech.edu/postbcd/doc/spice.pdf} 

	 University Press

\bibitem[\protect\citeauthoryear{Tabur \& Monard}{2003}]{tabur} Tabur, V.,
         Monard, L. A. G., 2003, IAUC~8134

\bibitem[\protect\citeauthoryear{Tielens}{2008}]{tielens} Tielens, A. G. G. M., 2008,
         ARAA, 46, 289

\bibitem[\protect\citeauthoryear{Werner et al.}{2003}]{spitzer} Werner M. W., et
         al., 2004, ApJS, 154, 1

\bibitem[\protect\citeauthoryear{Wooden, Woodward \& Harker}{2004}]{wooden}
        Wooden, D. H., Woodward, C. E., Harker, D. E., 2004, ApJ, 612, L77


%
%
%
%
%
%
%
%


\end{thebibliography}
\end{document}